\documentclass[journal]{vgtc}                       

\onlineid{1161}

\vgtccategory{Research}

\vgtcpapertype{Evaluation}

\title{2D, 2.5D, or 3D? An Exploratory Study on\\ Multilayer Network Visualisations in Virtual Reality}

\author{%
  \authororcid{Stefan P.\ Feyer}{0009-0004-8574-0741},
  \authororcid{Bruno Pinaud}{0000-0003-4814-3273},
  \authororcid{Stephen Kobourov}{0000-0002-0477-2724},
  \authororcid{Nicolas Brich}{0000-0003-3175-0464},
  \authororcid{Michael Krone}{0000-0002-1445-7568},\\
  \authororcid{Andreas Kerren}{0000-0002-0519-2537},
  \authororcid{Michael Behrisch}{0000-0002-1102-103X},
  \authororcid{Falk Schreiber}{0000-0002-9307-3254}, and
  \authororcid{Karsten Klein}{0000-0002-8345-5806}  
}

\authorfooter{
    \item 
        Stefan P. Feyer and Karsten Klein are with the University of Konstanz, Life Science Informatics.
        E-mail: \{stefan.feyer \textbar\ karsten.klein\}@uni-konstanz.de.
    \item 
      Bruno Pinaud is with Univ. Bordeaux, CNRS, Bordeaux INP, LaBRI, UMR 5800, France.
      E-mail: bruno.pinaud@u-bordeaux.fr.
    \item 
        Stephen Kobourov is with the University of Arizona.
        E-mail: kobourov@cs.arizona.edu.
    \item 
        Nicolas Brich is with University of Tübingen, Germany.
        E-mail: nicolas.brich@uni-tuebingen.de.
    \item 
        Michael Krone is with University of Tübingen, Germany and with New York University, USA.
        E-mail: michael.krone@uni-tuebingen.de.
    \item 
        Andreas Kerren is with Linköping University and Linnaeus University, Sweden.
        E-mail: andreas.kerren\{@liu.se, @lnu.se\}.
    \item 
        Michael Behrisch is with Utrecht University, NL.
        E-mail: m.behrisch@uu.nl.
    \item 
        Falk Schreiber is with University of Konstanz and Monash University.
        E-mail: falk.schreiber@uni-konstanz.de.
}

\abstract{Relational information between different types of entities is often modelled by a multilayer network (MLN) -- a network with subnetworks represented by layers. The layers of an MLN can be arranged in different ways in a visual representation, however, the impact of the arrangement on the readability of the network is an open question. Therefore, we studied this impact for several commonly occurring tasks related to MLN analysis. Additionally, layer arrangements with a dimensionality beyond 2D, which are common in this scenario, motivate the use of stereoscopic displays. We ran a human subject study utilising a Virtual Reality headset to evaluate 2D, 2.5D, and 3D layer arrangements. The study employs six analysis tasks that cover the spectrum of an MLN task taxonomy, from path finding and pattern identification to comparisons between and across layers. We found no clear overall winner. However, we explore the task-to-arrangement space and derive empirical-based recommendations on the effective use of 2D, 2.5D, and 3D layer arrangements for MLNs.}

\keywords{Network, Guidelines, VisDesign, HumanQuant, CompSystems.}

\teaser{
  \centering
  \includegraphics[width=\linewidth, trim={0 0 0 9mm},clip]{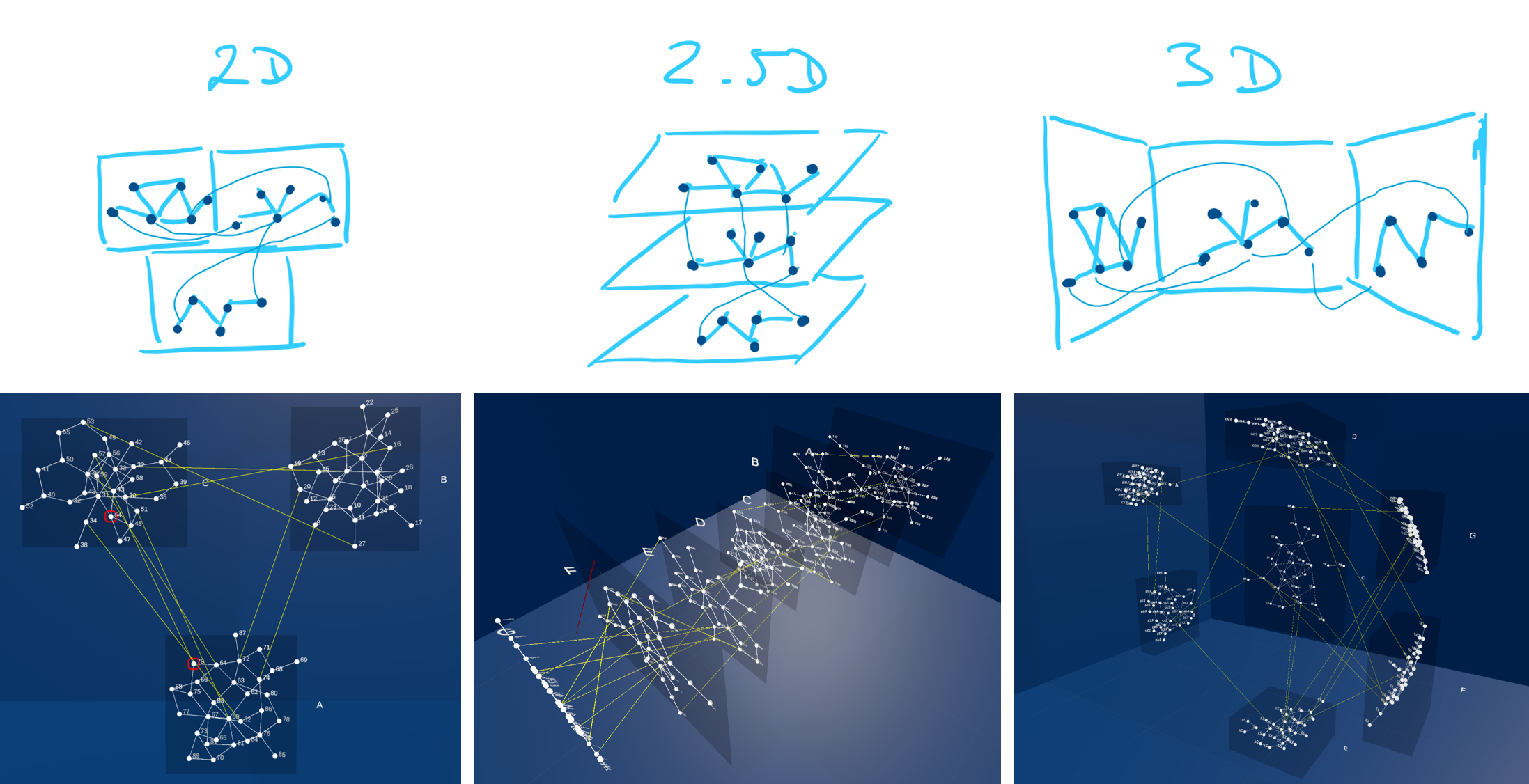}
  \caption{Examples of layer arrangements evaluated in this study for multilayer network visualisation. 
  Top: a hand-drawn sketch of small graphs (3 layers) from an early design draft. Bottom: examples of visualisations in a virtual reality setting used during the experiment. Note that 2D and 2.5D contains flat 2D areas for layers, whereas 3D is also for each layer a 3D representation. 2D shows a small graph (3 layers) whereas 2.5D and 3D show larger graphs (7 layers).}\label{fig:teaser}
}

\usepackage{mathptmx}                  

\usepackage{amsmath} 

\newcommand{\cit}[1]{``\textit{#1}''}

\begin{document}



\firstsection{Introduction}
\maketitle
In a variety of applications, we use networks with different types of entities and relations to model our data holistically. Knowledge graphs, for instance, combine data from diverse knowledge bases to express relations between genes, drugs, diseases, and pathogens in biomedical applications~\cite{NICHOLSON20201414} and, more broadly speaking, allow for inference of previously undiscovered patterns~\cite{Walsh20,Zheng20,Waagmeester20,li_kg4vis_2021}. In general, entities and relations of one type form 
layers
within the encompassing network, and the analysis of these networks requires taking the multilayer structure explicitly into account.
Specifically, each node in these so-called multilayer networks (MLNs) needs to be scrutinised concerning the layer it belongs to, i.e., interrelations within its layer 
and also the entangled effects between distinct layers.
We use the standard definition of an MLN from \cite{MLBBook21} inspired by \cite{Kivela2014}:
``In order to define a multilayer graph in a more formal manner, we begin with the definition of a standard graph.
A standard graph is often described by a tuple $G=(V,E)$ where $V$ defines a set of vertices and $E$ defines a set of edges (vertex pairs), such that $E \subseteq V \times V$.
An intuitive definition of a multilayer network first consists of specifying which layers the network nodes belong to.
Given a set of layers $L$, with an individual layer being defined as $l \in L$, and given that we allow a node $v \in V$ to be part of some layers and not others, we may consider nodes in a multilayer graph as pairs $V_M \subseteq V \times L$.
Edges $E_M \subseteq V_M \times V_M$ then connect pairs $(v, l), (v', l')$.''
Many different visual analysis approaches for MLNs have been proposed~\cite{McGee19,MLBBook21,Vehlow17,bianconi_multilayer_2018,Kivela2014,interdonato_multilayer_2020}. 
We claim that such analysis is significantly impacted by the layer arrangements and the complexity of the networks.
Certain arrangements and network complexities facilitate the interpretation of MLNs. 
In this work, we study the impact of layer arrangements and MLN complexities on MLN visual analysis.

Several metaphors for layer arrangements of MLNs have been proposed, including 2D, 2.5D and 3D~\cite{Berger19,cuenca_vertigo_2021,hammoud_multilayer_2020,Vehlow17,MLBBook21,dickison_magnani_rossi_2016} (\cref{fig:teaser}). Layer arrangements with a dimensionality beyond 2D
motivate the question: do immersive environments supporting stereoscopic 3D (S3D) help in solving MLN-based tasks? 
These environments are not only capable of providing S3D visualisations but also provide movement tracking and up to six degrees of freedom, which can support new ways of investigating and interacting with visual representations. The general investigation of the value of immersive visualisations has gained substantial momentum in recent years~\cite{kraus2021value}, and there is evidence of the usefulness for abstract data visualisation~\cite{kraus2022immersive}.
However, none of the studies has investigated how modern immersive environments affect the analysis of MLN. 
Further, there is a lack of empirical evidence regarding low-level task performance on MLN visualisations~\cite{McGee19,MLBBook21,pohl2019human}. This work aims to close this gap in the literature.

While several aspects of the encoding might have an effect on perception and task performance, we expect the MLN dimensionality to have a fundamental impact and should inform subsequent visualisation mappings, the interaction design, as well as the decision on the use of immersive environments. 
Consequently, the study presented in this paper focuses on the dimensionality- and layer-arrangement aspects of MLN analysis.
We compare three MLN representation settings with 2D, 2.5D and 3D metaphors (\cref{fig:teaser}), derived from McGee et al.~\cite{MLBBook21}, thereby covering established standard MLN representations, as well as the arrangements most suited for modern immersive environments. 

We tackle the question of the effectiveness of 2D, 2.5D, and 3D layer arrangements for node-link visualisations in stereoscopic virtual reality (VR) settings. 
In 2D, the layers are located on a plane (\cref{fig:teaser}, left), in 2.5D the layers depict stacked planes (\cref{fig:teaser}, middle), and in 3D the layers are arranged in 3D space on a hemisphere (\cref{fig:teaser}, right). 
To investigate the differences, we experimented with participants who had to solve six tasks derived from the MLN analysis task taxonomy proposed by McGee et al.~\cite{MLBBook21}. 
Our study contributes empirical evidence about task-solving performance concerning layer arrangements on two network complexities (3 and 7 layers).
The different visualisations and the process of our study are shown in the accompanied video.
Overall, the primary contributions of our work are:
    \begin{enumerate}[topsep=0pt,itemsep=-1ex,partopsep=1ex,parsep=1ex,leftmargin=1.5em,labelwidth=1.5em]
        \item a controlled, stereoscopic VR-based human subject study   investigating the effectiveness of 2D, 2.5D, and 3D layer arrangements for node-link visualisations on six distinct MLN analysis tasks;    
        \item a structured analysis of the MLN task-to-layer arrangement space along with human subject study-derived recommendations on the effective use of visual metaphors, layer arrangements, 
    and perceptual considerations;    
        \item the networks used in this experiment along with the code to generate them is offered to the community to facilitate future research.
    \end{enumerate}
    
The rest of this paper is organised as follows. \cref{sec:related} presents related work on MLNs and network visualisation in VR.
\cref{sec:design} describes the experimental design, task description, and the chosen methodology. 
\cref{sec:results} highlights the results of the study and presents qualitative feedback given by the participants.
This is followed in~\cref{sec:discussion} by a discussion on these results, and in \cref{sec:limitations}, we discuss the inherent limitations of every human subject experiment before concluding in \cref{sec:conclusion}.

\section{Related Work}\label{sec:related}
MLNs are important to model complex relations in a variety of scientific disciplines and domains, such as biology and bioinformatics~\cite{zhou_omicsnet_2018,liu_paintomics_2022,santos_knowledge_2022}, epidemiology~\cite{estrada_covid-19_2020}, social sciences and digital humanities~\cite{becker_multiplex_2020,bornhofen_exploring_2020}, economics~\cite{bardoscia_physics_2021}, or civil engineering~\cite{baggag_resilience_2018}.
The interactive visualisation of such networks can help end-users to better understand and analyse this kind of data, especially the relationships between different network entities such as nodes and/or layers.
A plethora of visualisation methods, task taxonomies, and domain-specific tools have been developed for the more general field of network visualisation and graph drawing; but for MLN visualisation the situation is still different, and related work in this area is by far more limited. Before we structure the original related work in this section according to dimensions such as available taxonomies, network visualisation, and human factors, we first highlight the existence of a number of surveys and state-of-the-art reports.

The book on multivariate network visualisation, edited by Kerren et al. in 2014~\cite{kerren2014multivariate}, includes a chapter written by Schreiber et al.~\cite{Schreiber2014} that is especially dedicated to multilayer visualisation.
It provides definitions of the terminology, application examples from various domains, and a discussion on the design space of visual designs for this specific kind of data. 
Kivelä et al.~\cite{Kivela2014} survey the existing literature on different types of MLNs, propose a terminology and a general framework for MLNs and discuss visualisation approaches for such networks. 
Bianconi~\cite{bianconi_multilayer_2018} discusses MLNs from a network science perspective.
In a work published later in 2017, Vehlow et al.~\cite{Vehlow17} present a literature survey about the visualisation of group structures in networks (those groups can be interpreted as layers in an MLN)
and also propose a taxonomy of corresponding visualisation techniques.
Similar to the above-mentioned edited book, Nobre et al.~\cite{Nobre2019} present a more recent state-of-the-art report on multivariate network visualisation and discuss MLNs as a special case.
Finally, the state-of-the-art report by McGee et al.~\cite{McGee19} specifically focuses on visualising MLNs and provides a comprehensive overview of this research area. In this paper, we mainly use and refer to the concepts, definitions, and terminology given in the recent book on MLNs by McGee et al.~\cite{MLBBook21}.

\subsection{Task Taxonomies for Visual Network Analysis}
\label{sec:relwork-tasktax}
A good visual design is always related to a set of tasks which should be supported by the final visualisation. Many of the above-mentioned surveys also include and discuss tasks for network visualisation and analysis. A general task taxonomy for network/graph visualisation is proposed by Lee et al.~\cite{lee_task_2006}, who extend the earlier, more general task taxonomy for information visualisation by Amar et al.~\cite{amar_low-level_2005}. On the highest level of their taxonomy, they propose the following task categories: topology-based, attribute-based, browsing, and overview tasks. While such general task taxonomies are an excellent starting point for the design and evaluation of network visualisations, it is worth noting that more specialised task taxonomies can be necessary to cater to the needs of a certain application domain. 
An example is the visualisation task taxonomy for biological pathways presented by Murray~\cite{murray2017taxonomy} that includes tailored tasks that require domain knowledge, such as the comparison of functionally similar pathways across species. 
In this paper, we use the task taxonomy, designed specifically for MLNs~\cite[Chap.~4]{MLBBook21}. There are four different task categories ranging from (A) entity connectivity across layers, (B) entity comparison across layers, (C) layer structure manipulations, and (D) layer comparisons. However, there is no specific category for defining an initial layering or considering specific domains. 

\subsection{Visualising Multilayer Networks in 2.5D and 3D}
\label{sec:relwork-vis2.5D-3D}
Van den Elzen and van Wijk present an approach for the exploration of multivariate networks based on selection and aggregation~\cite{vdElzen14}.
While node-link representations are dominating, there are also approaches that make use of matrix representations.
An example of this is the approach by Horak et al.~\cite{horak_responsive_2021}.
Berger et al. investigate the relation between structure and attributes~\cite{Berger19} (matrix visualisation). 
Approaches that make use of 2.5D and 3D network arrangements have been proposed in classical graph drawing (i.e., aiming at 2D projections), often by extending established layout methods into the third dimensions~\cite{Hong07,BrandesDwyerSchreiber+2004+11+26,de_domenico_muxviz_2015}.
Regarding the arrangement of layers or similar concepts, related work discusses how cluster graphs are positioned in space; especially for the 3D case Roberts et al.~\cite{Roberts21CGVC} advocate for adding further views, instead of just using a single 3D visualisation. 
While a number of related network visualisation approaches feature 2.5/3D visualisation~\cite{zhou_omicsnet_2018}, they are still less common compared to traditional 2D approaches.
In their systematic review, the authors of the survey~\cite{McGee19} point out that 3D visualisation has no scientific basis outside of stereo-vision/VR \cite{ware_visualizing_2008, van_kreveld_visual_2012}. We agree that the discussion on the usefulness of 3D network visualisation is ongoing for many years, but there are arguments which support the visual display of abstract data and 3D spatial information together as stated in the position paper by Kerren and Schreiber~\cite{KerrenS14}.

\subsection{Network Visualisation in Virtual Reality}
\label{sec:relwork-vr}
Immersive analytics is growing in popularity, thanks to the availability of affordable and mature head-mounted displays (HMD) for virtual and augmented reality (VR/AR) as well as mature software platforms such as Unity or Unreal Engine~\cite{unity}.
Visual network analysis has been proposed as a potential use case for such environments~\cite{cordeil_immersive_2017}.
The stereoscopic display capabilities of VR HMDs paired with the possibilities for natural interaction are supposed to remedy the usual drawbacks of three-dimensional visualisations of abstract data (e.g., occlusion or perspective distortion).
Recently, Korkut and Surer~\cite{korkut_visualization_2023} presented an extensive survey of visualisations in VR.
They concluded that advances in immersive visualisations often do not focus on building theoretical backgrounds and that the constant technological changes require repeated studies to ensure that the theories are still valid. This is very much in line with the focus of our paper.

While not concerned with network visualisation itself, several studies investigate relevant phenomena pertaining to the layout navigation and linking of visualisations in 3D and VR scenarios.
Shoemake et al.~\cite{shoemake} describe a technique for navigation in 3D environment.
Liu et al.~\cite{liudesign} study the different spacial arrangements of small multiples in VR.
Prozeau et al.~\cite{prouzeauvisual} move closer to the network visualisation space by laying out links between coordinated views in a 2.5D VR scenario.
Similarly, Yang et al.~\cite{yangorigindestination} investigate how to best link points on a map in an immersive visualisation scenario.
Butscher et al.~\cite{butscherclusters} apply the concept of parallel coordinate plots to VR which offers challenges similar to 2.5D graph visualisations in VR. 
MetNetVR by Yang et al.~\cite{yang_hierarchical_2006} is an example of an early network visualisation in a CAVE VR environment. It offers visualisation of biological pathways with level-of-detail functionality. However, the authors do not focus on optimising the layout for this purpose but used a standard three-dimensional force-directed layout algorithm.
Min\-Omics by Maes et al.~\cite{maes_minomics_2018} offers a 3D visualisation of proteomics networks in VR based on UnityMol~\cite{doutreligne_unitymol:_2014}. While the 3D layout is not explicitly discussed, it again seems to use a force-directed layout algorithm.
More recently, Pirch et al.~\cite{pirch_vrnetzer_2021} developed VRNetzer, another VR application for visualising molecular biological networks.
Although it includes different layout options, these are not in the focus of the paper, and the authors do not evaluate their effectiveness.
Yang et al.~\cite{yangembodied} study navigation methods in VR in general, while Sorger et al.~\cite{sorgerimmersive} and Drogemuller et al.~\cite{drogemullerexamining} apply this to 3D graph visualisation in VR. 
Kwon et al.~\cite{kwonstudy}  investigate the layout and rendering of network visualisations, as well as interactions, in VR.  
In a subsequent work, Sorger et al.~\cite{sorgeregocentric} also investigate the influence of applying different variations of egocentrism in network layout techniques in VR.
Joos et al.~\cite{joos_visual_2022} present a study on the visual comparison of weighted graphs in VR.
However, similar to the above-mentioned VR graph visualisations, these studies do not consider multilayer networks.
One of the few currently available tools for MLN visualisation in VR is MNET-VR~\cite{MNET} that is still under active development. 
Our work aims to fill the gap in the literature concerning the effectiveness of different layouts for the visual analysis of MLNs.

\subsection{Evaluation and Human Factors}
\label{sec:eval-humanfactors}
Pohl and Kerren~\cite{pohl2019human} discuss human factors in the context of MLN visualisation and highlight the most important related studies performed in the fields of human-computer interaction and cognitive psychology that might be helpful for future designers of MLN visualisations. 
The authors also propose a number of first (perhaps tentative) design guidelines for MLN visualisations and identify research challenges when it comes to their evaluation. 
Other studies, such as the one performed by Kotlarek et al.~\cite{kotlarek_study_2020} investigate the performance of VR in the context of general network visualisation, to the best of our knowledge, there is no existing work focusing on MLN visualisation in VR. 
In consequence, there are no conclusive and robust results for the utility of VR for MLN visualisation yet. 
Thus, we took the paper by Pohl and Kerren~\cite{pohl2019human} as motivation for the study presented in this paper (for the visual design in VR and for the methodology of the study itself) and especially considered their recommendations with respect to cognitive load theory, i.e., we emphasise that the \cit{cognitive load for remembering relationships while solving tasks does not become too heavy} and more \cit{complex tasks should be used more often to evaluate graph visualisations}, see \cref{sec:design}. Note, that our proposed study solely 
regards network representations in the shape of node-link diagrams and no matrix, hybrid, or other representations, such as discussed by Roberts et al.~\cite{Roberts2014} or compared by Okoe et al.~\cite{Okoe19}.

\section{Experiment Design}\label{sec:design}
The main focus of our work is to study the impact of layer arrangement on MLN visual analysis. We decided, therefore, to discern layer \textit{arrangements}, i.e., how MLN layers are depicted to the participants: 2D (\cref{fig:teaser}, left), 2.5D (stacked layers, \cref{fig:teaser}, middle), and 3D (\cref{fig:teaser}, right). We chose the last two arrangements as they can benefit from stereoscopic 3D and changes in the viewing perspective. 

For the number of layers, we decided to start with the smallest non-trivial number, i.e.,~three, having at least one intermediate layer. 
To measure the potential effects of scale, while keeping the number of different conditions low, we decided to have one additional layer setup that we expect to increase the difficulty significantly without approaching a perceptual or cognitive limit. After several initial pilot experiments including a visual inspection by the authors, we settled for seven layers as the higher complexity condition.
For ease of reading,
we refer to networks with 3 layers as ``small'' and to networks with 7 layers as ``large'', while acknowledging that in general networks with 7 layers are not considered large in real-life use cases.

In all experiments, we made use of a VR environment as it allows a change of perspective through head-tracking and body movement and provides stereoscopic 3D vision support in particular for the 2.5D and 3D settings. 
To minimise confounding factors based on changes in the environment or based on interaction operations, we conducted all experiments in the same VR environment and with head/body movement and pointing being the only allowed interaction sources. Furthermore, we decided to settle on node-link visualisations as they are the dominant idiom for network visualisation in practice and allow us to keep the same visual encoding across the three layer arrangements. 

As explained in \cref{sec:related}, to achieve good coverage of task classes for MLNs, we used the MLN task taxonomy given by McGee et al.~\cite{MLBBook21}. 
They identified four main types of tasks, A) entity connectivity across layers, B) entity comparison across layers, C) layer structure manipulations, and D) layer comparisons. 
For our goal of investigating fundamentals of MLN readability for three layer arrangements, we propose tasks based on categories A, B, and D. Tasks of type C seem to not fit properly, as they would mainly target interactions to change the network structure. As our focus is on exploration, we want to focus on the fundamentals, that is the support for readability and understanding in 2D, 2.5D, and 3D layer arrangements.

Several characteristics might influence the experiment: task difficulty and performance results, notably the network size (number of nodes and edges) and the number of layers.
As our main aim is a comparison between layer arrangements, we decided not to include further variations of these characteristics beyond the two layer numbers as conditions. On the one hand, this guarantees a manageable extent of study conditions and on the other hand, provides a starting point for further investigations regarding the influence of these characteristics. 
Thus, for the tasks that cover entity connectivity (Category~A) and entity comparison (Category~D), we use the same size for the MLN and only vary the number of layers as the main impact factor of task complexity. For the tasks which cover layer comparison, we vary the density of the MLN across different layers. 

\subsection{Network generation, layouts, and arrangements}
\label{sub:layout}

To have full control of the study conditions 
we decided to generate our networks so that we can provide a comprehensive cross-section of network characteristics.
We created a large number of random networks (36k) and then filtered them out based on the desired feature ranges (12k).
These feature ranges are supposed to support comparability and also feasibility of the tasks within our settings. They were chosen based on previous experience and tests with the VR environment before the study. 
In order to perform the generation, we created a generator that allows us to tune parameters and select different models for network generation and layouts. For the implementation, we used the \textit{Open Graph algorithms and Data structures} (OGDF) framework~\cite{chimani2013open}. 

Our procedure for MLN generation is as follows:

\begin{enumerate}[topsep=0pt,itemsep=-1ex,partopsep=1ex,parsep=1ex,leftmargin=1.5em,labelwidth=1.5em]
    \item for a number $l$ of layers, we randomly generate a network for each layer with $n$ nodes and $m$ edges using a generator model \textit{gmod}; 
\item the layers are connected
using a network generator model \textit{imod};
\item each edge in the network created using model \textit{imod} is a connection between nodes in different layers, with 

a total of $im$ such edges, connecting random pairs of nodes in the corresponding layers.
\end{enumerate}
Parameters $l$, $n$, $m$ and, $im$ can be freely chosen. For our study, we used $l=3\text{ or }7$; $n = 30$; $m = 45$ and $im = l*ilef$ where $ilef$ is a factor that is set to $3$ in order to have a number of edges between layers linear in the number of layers but on the lower end of the density spectrum. 
For the generator models \textit{gmod} and \textit{imod}, any model implemented with/in OGDF can be chosen. We used the random simple connected graph generator of OGDF for \textit{imod}, where we extended the implementation for the layer connection to allow multiple connections between layers. For \textit{gmod}, we used generators as follows:
for the generation of MLNs, we used the following options (see \cref{sub:tasks} for the task descriptions):
\begin{itemize}[topsep=0pt,itemsep=-1ex,partopsep=1ex,parsep=1ex,leftmargin=1.5em,labelwidth=1.5em]
    \item Connected networks with equal density on each layer - using the OGDF \textit{simple connected graph generator}.
    \item Connected networks with varying density on each layer - by defining a span of $m-10$ to $m+10$ edges and by randomly picking a number of edges for each layer out of that span. We check for the resulting span of edge numbers for each generated network and for the gap between the highest and second highest density to judge the difficulty of the ``highest layer density'' task.
    \item Non-connected networks, where we randomly distribute the occurrence of $2*l$ connected components to the layers for the ``connected components'' task. Each connected component is created using the \textit{simple connected graph generator}.
    \item Connected networks that contain a $K_{2,3}$ pattern - using the OGDF \textit{simple connected graph generator} and by testing the occurrence of such a pattern for the ``connectivity pattern'' task. Either one or two layers contain the pattern in our networks.
\end{itemize}

Our procedure for the layout and arrangement is as follows: Let $MG$ be the input MLN, and $GL$ be the graph that is created from an empty graph by adding a node for each layer of $MG$ and an edge for each edge between layers of $MG$. 
That is, the MLN is interpreted as a simplified graph in which each layer is a node and the edges between layers induce edges in $GL$. First, we calculate a layout for each layer graph in $MG$ using a layout method $lmeth$. 
Then, an arrangement for $MG$ is created as follows.
For the 2D layer arrangement, the bounding boxes of the layer layouts are used to derive the node sizes in $GL$, and a force-directed layout is calculated to achieve an arrangement of the layers in 2D.
The layer graphs are rotated to minimise the square of the lengths of the edges between layers.
For the 2.5D layer arrangement, the layer layouts are stacked horizontally. The order is derived from the order of the vertices in one direction of a force-directed layout of $GL$.
For the 3D layer arrangement, we use stereographic projection of the 2D layout on a hemisphere.

We used carefully generated synthetic data for the experiments so that we could control as many of the variables involved as possible (e.g., density of the networks, node degree distribution, size of largest clique, etc.). 
The generated networks, while exhibiting a certain range of features, cover  only a small subspace of the potential network design space. Nevertheless, our network generator (available in supplementary materials) can be used for future experiments.

\subsection{Tasks description}\label{sub:tasks}
Our study tasks are derived from the task taxonomy given by McGee et al.~\cite{MLBBook21}.
In total, each participant had to complete six tasks named T1, T2, \ldots, T6.
T1/T2 relate to task category A (entity connectivity across layers). 
T3/T4 map to category B (entity comparison across layers), 
and T5/6 link to category~D (layer comparisons).
Below, the tasks are described and illustrated in \cref{fig:allTasks}. We list the exact question asked to the participants in the headings below.

{\setlength{\tabcolsep}{2pt}
\begin{figure*}[ht!]
\centering

\begin{tabular}{c|c|c}

        \includegraphics[width=0.3\textwidth]{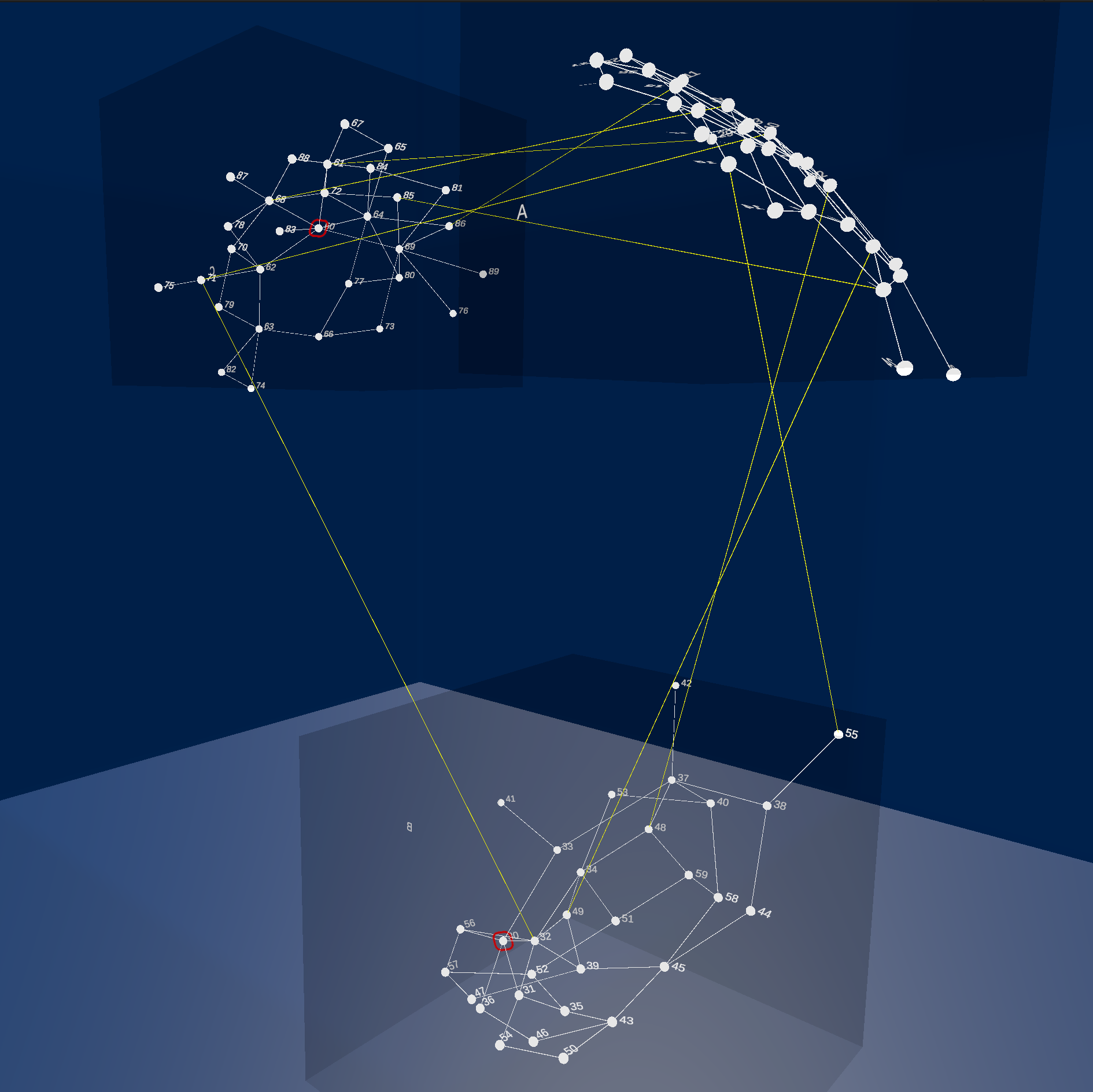}&
        \includegraphics[width=0.3\textwidth]{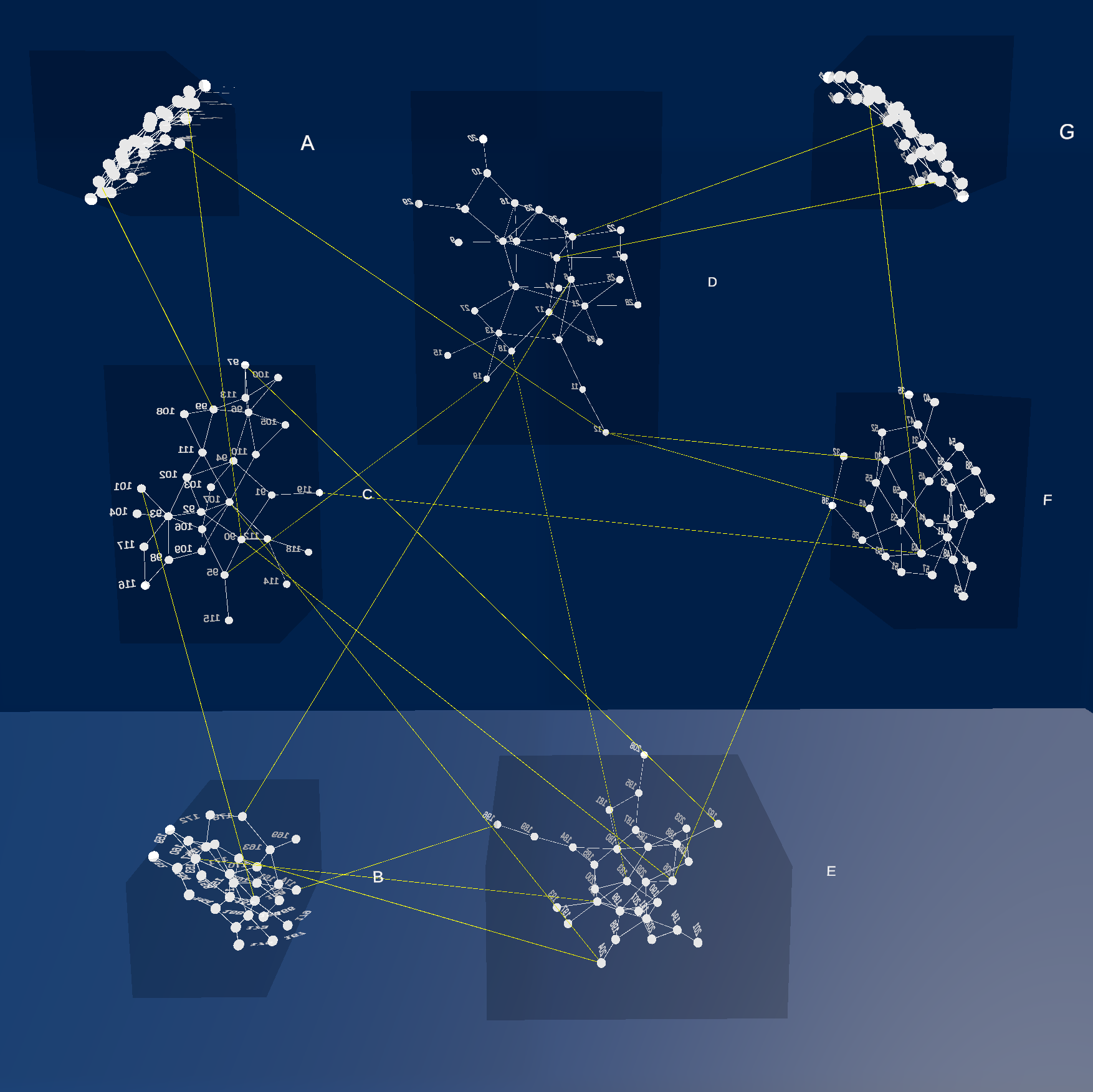}&
        \includegraphics[width=0.3\textwidth]{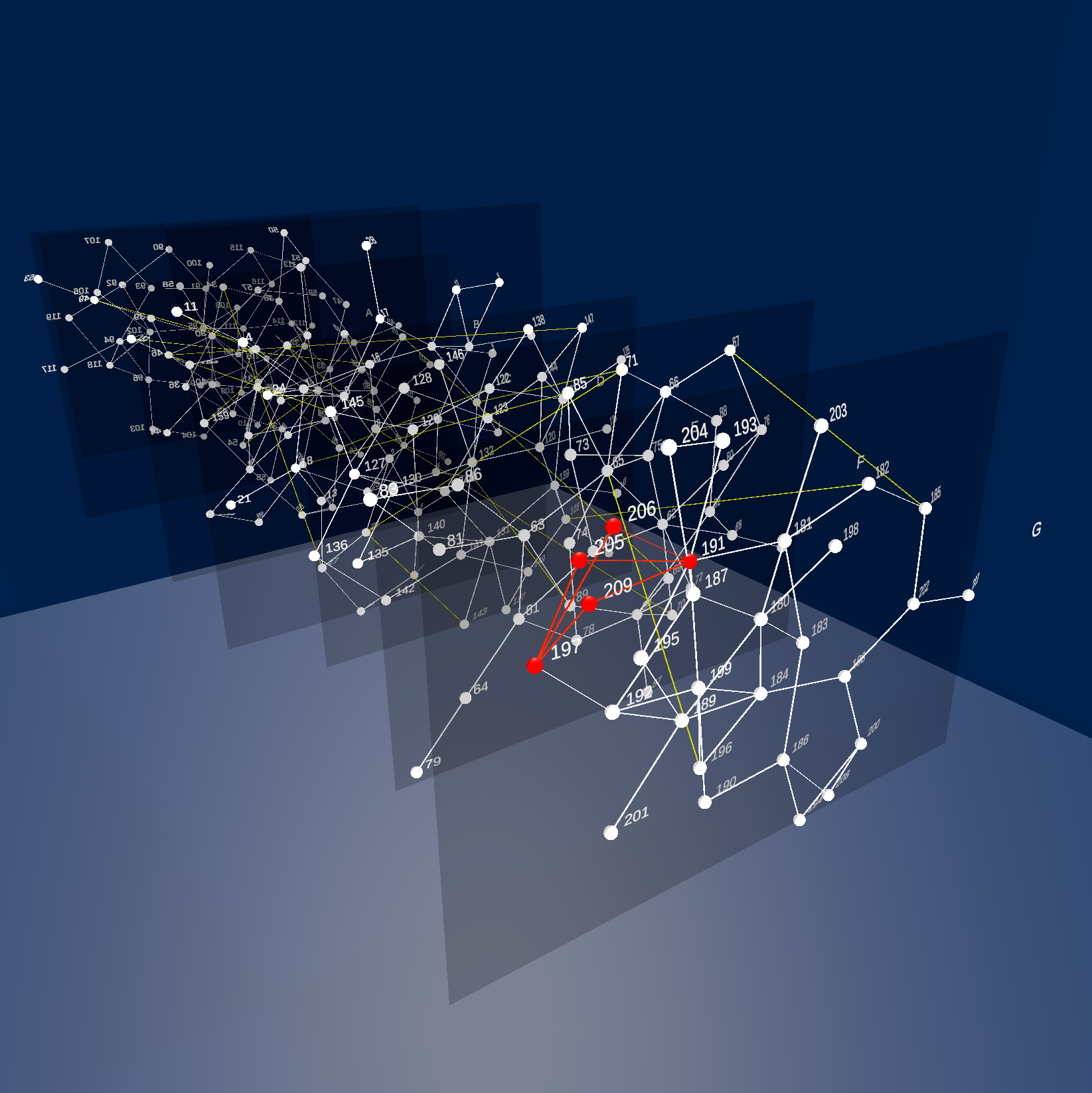}\\    
        \textbf{T1 (3D, small network)} & \textbf{T2 (3D, large network)} & \textbf{T3 (2.5D, large network)} \\
        \shortstack{How long is a shortest path between\\ the two red highlighted nodes?}&
        \shortstack{Which layer is connected\\ to most other layers?}&
        \shortstack{Does the same connectivity pattern as the\\ highlighted one also appear on other layers?}\\

        \includegraphics[width=0.3\textwidth]{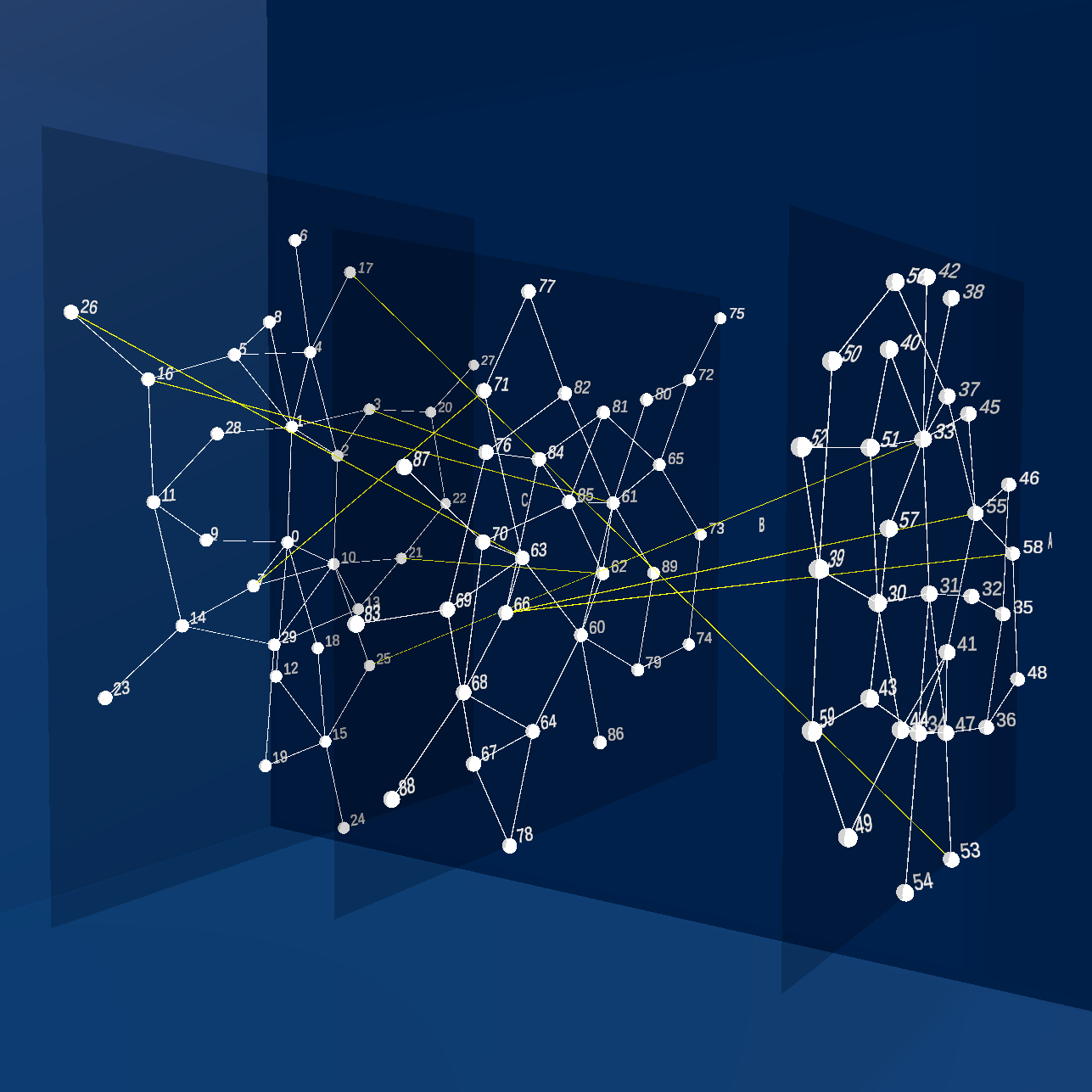}&
        \includegraphics[width=0.3\textwidth]{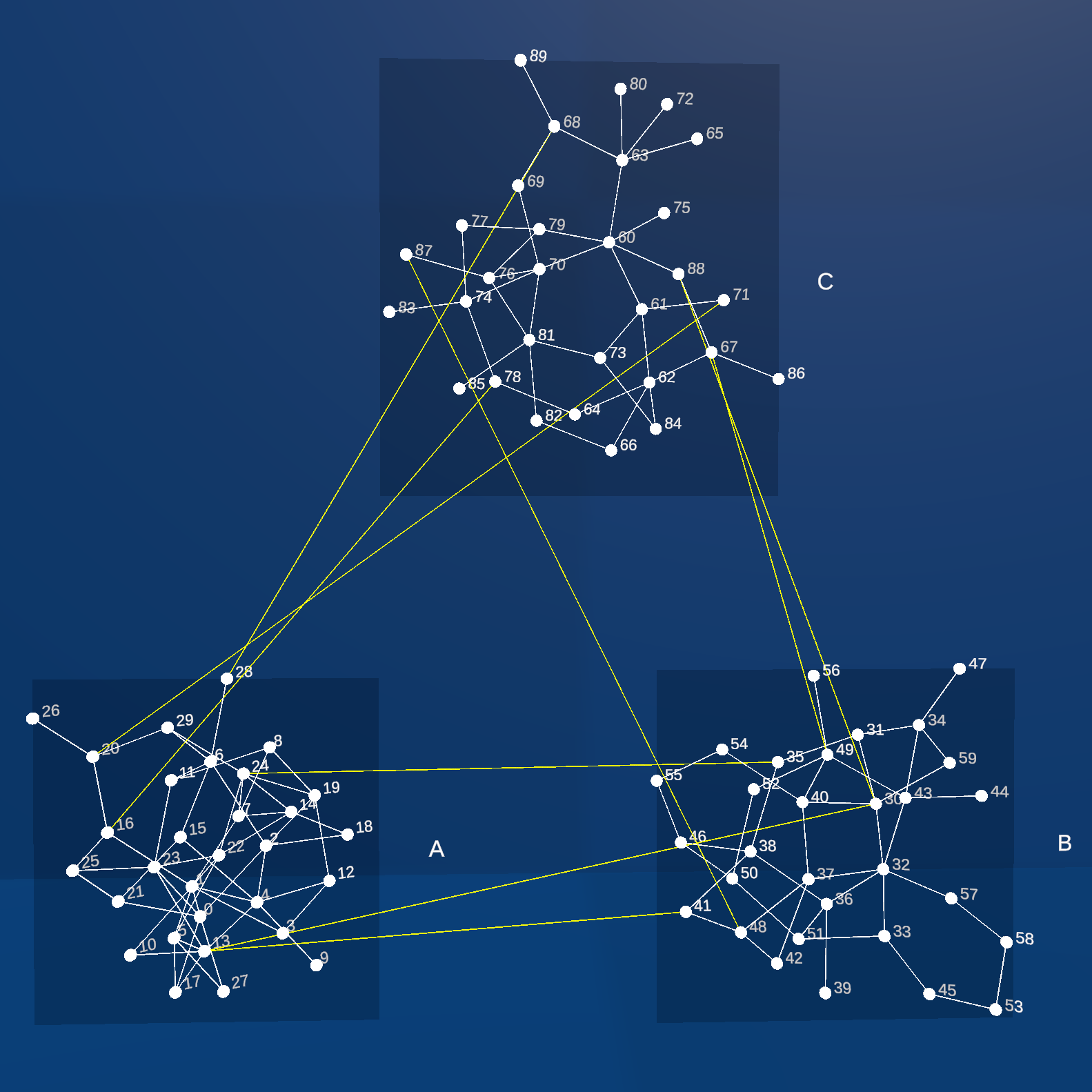}&
        \includegraphics[width=0.3\textwidth]{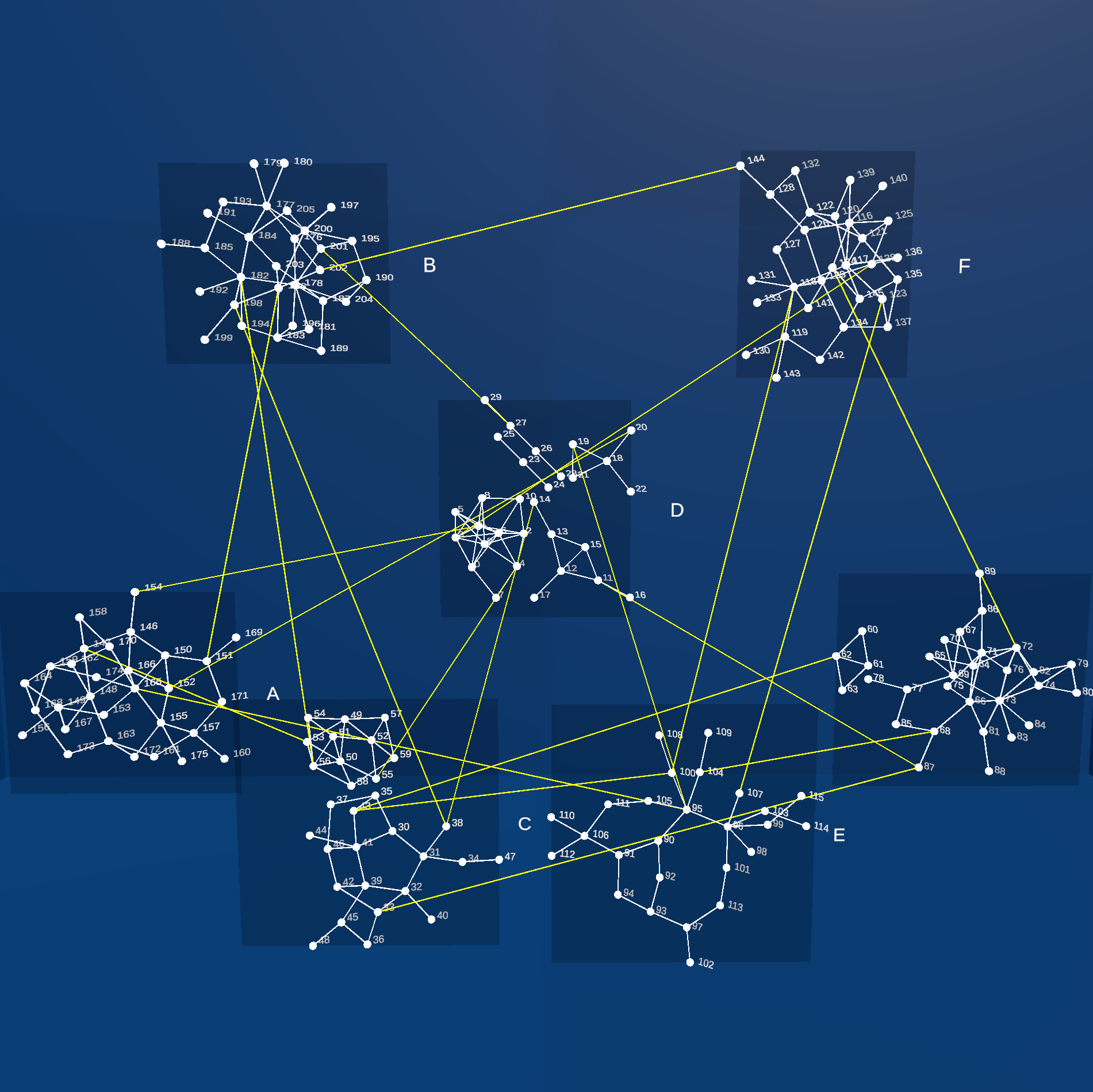}\\      
    \textbf{T4 (2.5D, small network)} & \textbf{T5 (2D, small network)} & \textbf{T6 (large network in 2D)} \\
      \shortstack{Which node has the highest\\ number of edges to other nodes?}&
      \shortstack{Which layer has the highest\\ amount of edges?}&
      \shortstack{Which layer has the \\most connected components?}\\
    \end{tabular}
        \caption{Illustration of the 6 tasks along with the exact question asked for different networks used during the experiment.}
        \label{fig:allTasks}
\end{figure*}}

\paragraph*{T1: How long is a shortest path between the two highlighted nodes?}
Two nodes, source and target, located on different layers are selected and highlighted with a red circle.
The path between the source and the target always passes through several layers. For better comparison of the tasks, the complexity is controlled by setting the path length between 3 and 5. The participants have to search for one (of potentially several) shortest-path alternatives. The length of the path is the number of edges on the path between the source and the target. The answer panel shows buttons for all possible answers (0 to 9). 

\begin{figure}[tb]
    \centering
    \includegraphics[width=0.8\columnwidth]{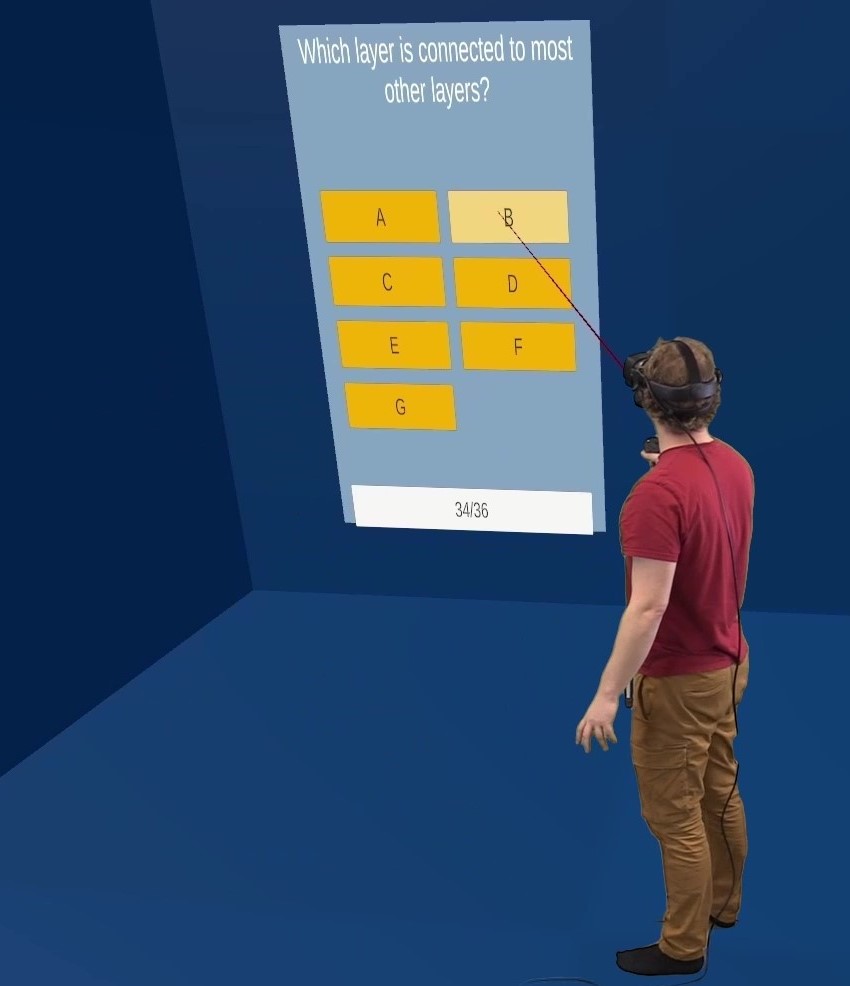}
    \caption{The panel on the wall of the VR environment used to guide the participants through the experiment and to provide answers.}
    \label{fig:panel}
\end{figure}

\paragraph*{T2: Which layer is connected to most other layers?}
The participant has to retrieve the layer with the most connections to other layers. The total number of edges between layers is 9 for small networks and 21 for large networks. The answer panel depicts the layer label (a capital letter) for all layers present in the current experiment step as individual buttons (\cref{fig:panel}).

\paragraph*{T3: Does the same connectivity pattern as the highlighted one also appear on other layers?}
On one layer, nodes and edges of a complete bipartite $K_{2,3}$ graph pattern are highlighted in red. 
The participants have to search for the exact pattern on other layers and this is a yes/no question.

The chance that there is another pattern is 50\%. We found out in several pilot studies that matching a connectivity pattern is a complex task. 
To ease the task, we decided to make our connectivity pattern more salient: a $K_{2,3}$ has 5 nodes, 3 of them with node degree 2 and 2 of them with node degree 3. We restricted the 2-degree nodes to not have any further edges to the rest of the network.
This results roughly in two main shapes of $K_{2,3}$: a crown like-shape and a diamond like-shape. Additionally, all nodes and edges of one $K_{2,3}$ are colour-coded in red. 

\paragraph*{T4: Which node has the highest number of edges to other nodes?}
The node degree describes the number of edges (between or inside layers) connected to a node.
The participants search for the node with the highest degree. Participants have to enter the chosen node label on the answer panel.

\paragraph*{T5: Which layer has the highest amount of edges?}
All layers have the same amount of nodes. Thus, the task is to find the layer with the highest number of edges. 
The difference in edges per layer is determined by randomly creating layer graphs with a uniform number of edges in a range of $\pm10$ edges around the standard number of edges per level ($m=45$). 
The minimum difference between the densest and second densest layer is at least 10 edges.
The answer is the label of the layer, all layer labels are displayed on the answer panel.

\paragraph*{T6: Which layer has the most connected components?}
The total number of connected components equals twice the number of layers. 
We use OGDF's ComponentSplitterLayout to create an arrangement of the components for each layer.
The participant has to determine the number of connected components per layer individually. The label of the layer with the highest number of connected components should be selected on the answer panel. 

\subsection{Experiment design and setup}\label{sub:setup}
Following the experimental design methodology of Purchase~\cite{Purchase:2012:EHCIP}, we created an exploratory within-participants experiment. 
Excluding the tutorial, the experiment is composed of the three-layer arrangements (2D, 2.5D, 3D) used with two types of network complexities (3~layers and 7~layers, a.k.a. small and large networks)
and six tasks previously presented (T1, \ldots, T6).
Consequently, participants had to answer $3 \times 2 \times 6 = 36$ questions within the study. The answers to each task are unique, and this fact is known by the participant.

The complexity levels and layer arrangements are counterbalanced in a Graeco-Latin square. 
We assume the learning effect among the tasks is low and further counteracted learning effects by randomising all tasks. 
To reduce a learning effect based on network repetition, a randomly chosen generated network was used for each task. 
In other words, no participant faced the same network twice. 

\subsubsection{VR Environment}\label{sub:vr}
The experiment was implemented with the game engine Unity~\cite{unity}. 
We chose a colourblind safe, consistent visual design for the study variants. Nodes were labelled with ascending numbers starting at 0. Layers were labelled with ascending capital letters starting at A. Edges between layers were displayed in yellow. Edges inside layers and nodes were depicted in white. The layers are drawn surrounded by a dark transparent grey bounding box (\cref{fig:allTasks}). Colours were chosen to provide good contrast between the network representation and the background.

In the VR environment, the participant was placed in a $5 \times 5$ metres room with dark blue walls. 
On one wall, a virtual answer panel was placed (\cref{fig:panel}).
The panel was the sole interaction surface and operated by a raycasting-based interaction approach for which we used a 3D-tracked \textit{Valve Index} VR controller. No participant had problems with this interaction method. For the experiment we used the VR head-mounted display \textit{Valve Index}.
The participant can navigate through the experiment through the panel: first, the question appears, and the participant has time to read the question. As soon as the participant clicks on \textit{start task}, the network appears and the participant starts to solve the task. After finding an answer, the participant clicks on \textit{done}, which leads to a screen on which the answer will be chosen. The network is not displayed anymore. After the answer is given by the participant, a confirmation question appears, which allows them to correct the given answer if needed. After confirming the answer, the next task appears on the panel. 

\subsubsection{Experiment protocol}
Participants started by signing a consent letter and completed a demographic questionnaire. After that, the participants followed a tutorial to learn MLN basics and to explain all tasks. Next, participants put on the VR headset. By performing a walk-around, they gained the confidence not to walk into a wall. 
In the subsequent training phase, the participants performed each task and practised using the panel to provide answers. The correct answer was shown on the panel during this phase. The training phase was also used to adjust the VR headset to a perfect fit. After the participant and the experiment conductor agree on a sufficient understanding of the tasks, the evaluation could really begin. 
During the study, participants were allowed to ask questions but did not get hints or indications pointing to the task results. After finishing all tasks, participants completed a survey to assess their subjective experiences. Lastly, they were compensated with 15€.

\section{Results}\label{sec:results}
We conducted our study at Utrecht University.
The participants were between 17 and 41 years old, with an experience with graphs ranging from none to very experienced, and an experience regarding VR from none to very experienced.
The counterbalancing of layer arrangements and complexity levels in a Greaeco-Latin square requires 12 participants ($3! \times 2!$); we aimed for two blocks (24 participants).
Overall, 28 participants took part in the experiment but, we could use only 22 results.
One participant could not wear the VR headset because of problems with glasses, another had field-of-view issued due to varifocal lenses, and two did not perform the tasks properly. 
Two more participants were excluded because of high error rates.
They did not understand the tasks or did not put enough commitment into the experiment.
Each participant's answers were analysed as follows:
\begin{itemize}[topsep=0pt,itemsep=-1ex,partopsep=1ex,parsep=1ex,leftmargin=1.5em,labelwidth=1.5em]
    \item The \textit{error rate} is the ratio of incorrect answers to the total.
\item The \textit{task completion time} (in seconds) refers to the time interval when participants work on the given task.
\item The physical \textit{movement performed} (in metres) is computed from the positions of the VR headset stored for every displayed frame. 
\end{itemize}
The VR headset position evolved when participants walked or rotated their head.
Overall, the experiment is physically engaging. On average, per participant, the measured completion time  is about $43.5$ minutes (std. dev. 10.5 min) and the amount of movement performed is about $320$m (std. dev. 75m). The preparation phase (headset setup, practice) lasted about 45 minutes.

We analyse the impact of the layer arrangements (2D, 2.5D, and 3D) for both network complexities (small and large) and the impact of the network complexities for each layer arrangement.
We employ pair-wise statistical tests when looking for statistically significant results. From a traditional standard significance level ($\alpha=0.05$), we need to apply a Bonferroni adjustment to take into account the multiple tests. 
For the layer arrangement analysis, we need $3\text{ tests} \times 2 \text{ complexities} = 6 \text{ tests}$.
For the network complexity analysis, we need $3$ tests. 
Thus, for each task and each measure, we perform $6+3=9$ pair-wise statistical tests.
As a consequence, for each test, we need to have $\alpha= 0.05/9= 0.0055$.
The error rate is computed from Boolean data (yes/no for each task). Thus, we use a Fisher’s exact test to determine whether a statistically significant relationship exists between two conditions. The null-hypothesis is there is no relationship between the conditions, i.e., knowing the performance of the participants on one condition (e.g., small networks 2D) does not provide information on the performance on another condition (e.g., large networks 2D).
To analyse completion time and movement performed, we use a Wilcoxon test, with a null-hypothesis that the two compared groups are equivalent.

The results for the 22 participants are presented in \cref{tab:resultsT1-T3,tab:resultsT4-T6} following the task taxonomy of McGee et al.~\cite{MLBBook21}. 
Blue (green) bars are for small (large) networks.
The error bars represent the 95\% confidence interval of the mean. Black lines between two layer arrangements indicate significant differences.
Black $\sqcap$-shaped lines indicate significant differences between the same arrangement of two network complexities. When $p\text{-values}<\alpha$, the values are in the figure captions. All $p\text{-values}$ are given in the supplemental material. 

{\setlength{\tabcolsep}{2pt}
\begin{table*}[!ht]
\centering
\begin{tabular}{|c|c|c|}
\hline
Task 1 & Task 2 &Task 3 \\
\hline
\multicolumn{3}{|c|}{Error rate}\\
\hline
\includegraphics[width=0.25\textwidth,trim={20 20 10 10},clip]{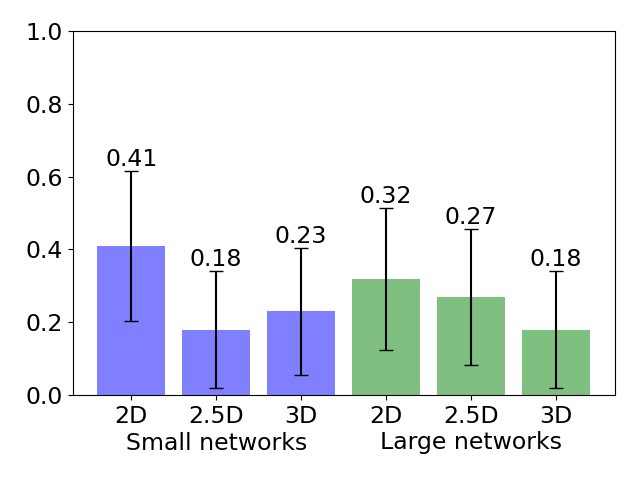}&
\includegraphics[width=0.25\textwidth,trim={20 20 10 10},clip]{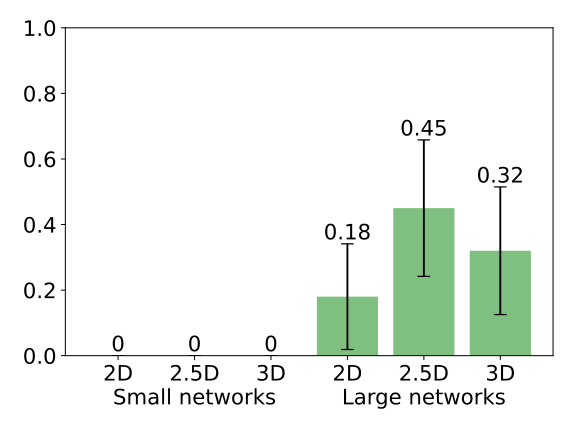}&
\includegraphics[width=0.25\textwidth,trim={20 20 10 10},clip]{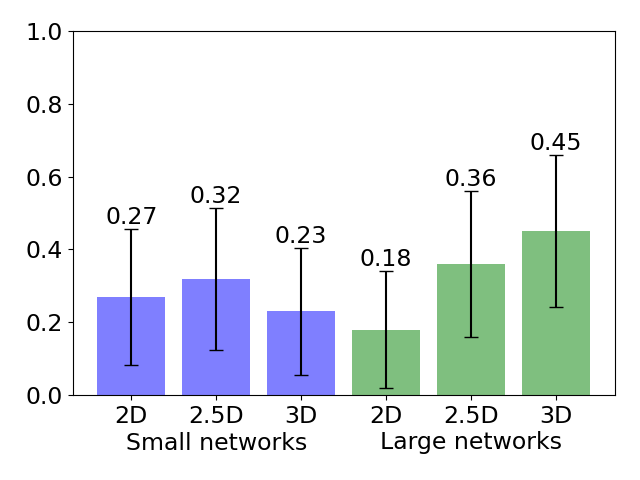}\\
\hline
\multicolumn{3}{|c|}{No test can be computed with T2 due to floor effects.}\\
\hline
\multicolumn{3}{|c|}{Completion Time}\\
\hline
\includegraphics[width=0.25\textwidth,trim={20 20 10 10},clip]{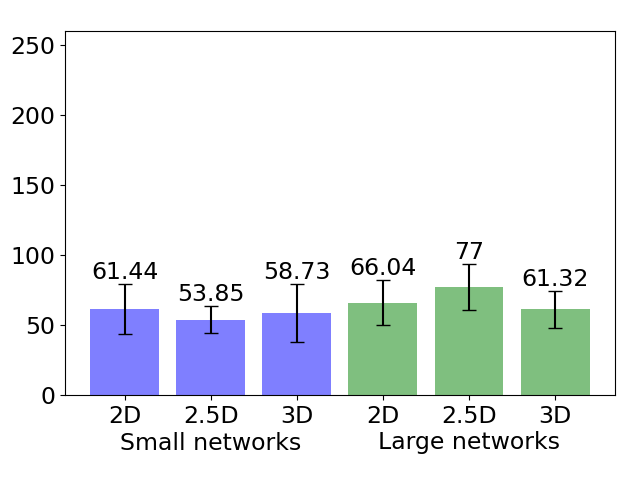}&
\includegraphics[width=0.25\textwidth,trim={20 20 10 10},clip]{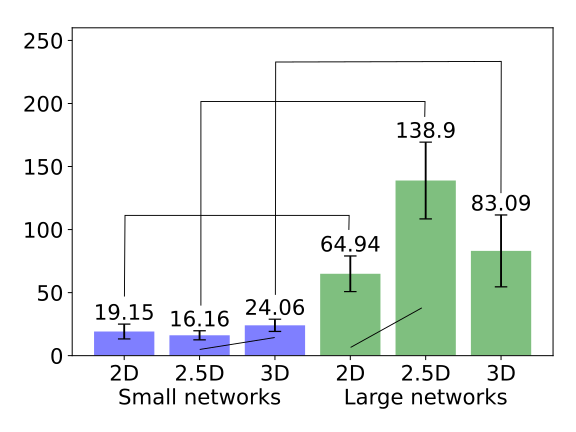}&
\includegraphics[width=0.25\textwidth,trim={20 20 10 10},clip]{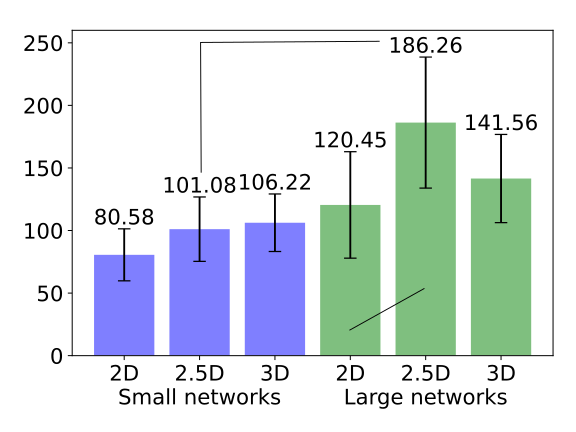}\\
\hline
\shortstack{We found no significant difference.}&
Significant $p\texttt{-values} \leq 0.001$.&
\shortstack{Significant $p\texttt{-values} \leq 0.003$.}\\
\hline
\multicolumn{3}{|c|}{Movement performed}\\
\hline
\includegraphics[width=0.25\textwidth,trim={20 20 10 10},clip]{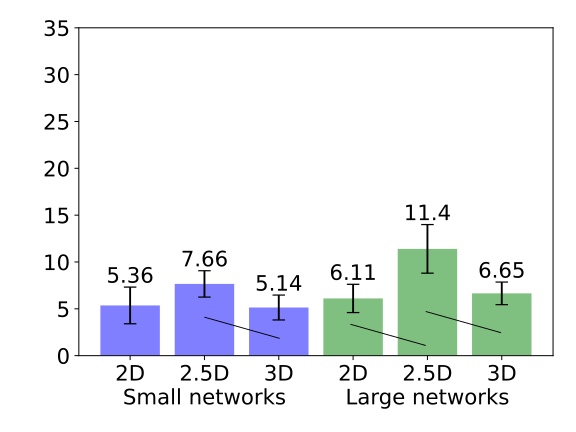}&
\includegraphics[width=0.25\textwidth,trim={20 20 10 10},clip]{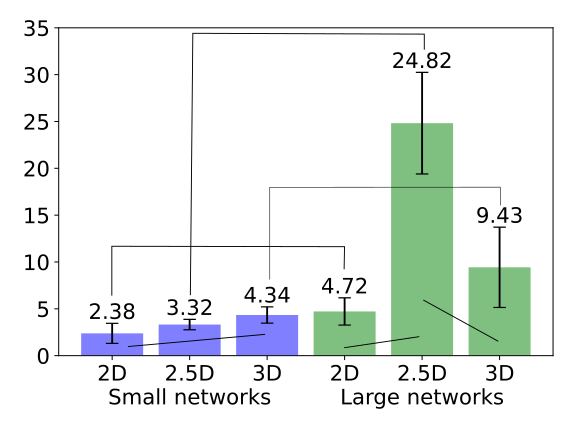}&
\includegraphics[width=0.25\textwidth,trim={20 20 10 10},clip]{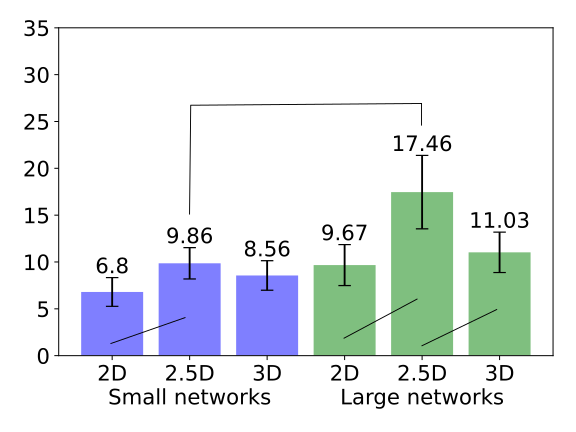}\\
\hline
Significant $p\texttt{-values} \leq 0.001$.&
Significant $p\texttt{-values} \leq 0.004$.&
Significant $p\texttt{-values} < 0.003$.\\
\hline

\end{tabular}

\captionof{figure}{Mean error rate, completion time and movement performed for T1, T2, and T3.}\label{tab:resultsT1-T3}
\end{table*}
}

\subsection{Entity connectivity across layers (T1, T2)}

The error rate in T1 (shortest path) reveals no relationship. However, for both complexity levels, the error rate is highest in 2D (0.41 for small networks and 0.32 for large networks). With small networks, the lowest error rate is in 2.5D (0.18) and in 3D for large networks (0.18). No high variability is present in the task completion time. The movement performed by the participants are higher in 2.5D (11.4m which is about 5 seconds more, compared to 2D and 3D).

Task T2 (degree of a layer) is affected by a floor effect for small networks (the {task} is too easy). Because we have no data for the error rate for small networks, the statistical tests cannot be computed (same for T5 and T6 below).
For large networks, the error rate is highest in 2.5D (0.45) and lowest in 2D (0.18). Notably, task completion time and  movement performed by the participants are {significantly} highest {in 2.5D (138.9s, 24.82m) and significantly} lowest in 2D (64.94s, 4.72m).

{\setlength{\tabcolsep}{2pt}
\begin{table*}[!ht]
\centering
\begin{tabular}{|c|c|c|}
\hline
Task 4 & Task 5 &Task 6 \\
\hline
\multicolumn{3}{|c|}{Error rate}\\
\hline
\includegraphics[width=0.25\textwidth,trim={20 20 10 10},clip]{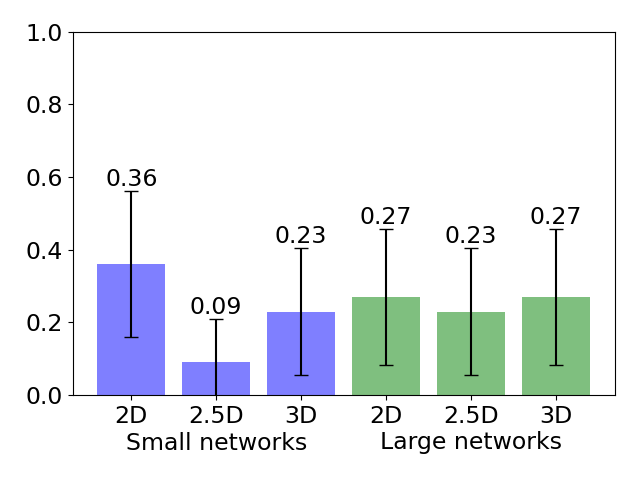}&
\includegraphics[width=0.25\textwidth,trim={20 20 10 10},clip]{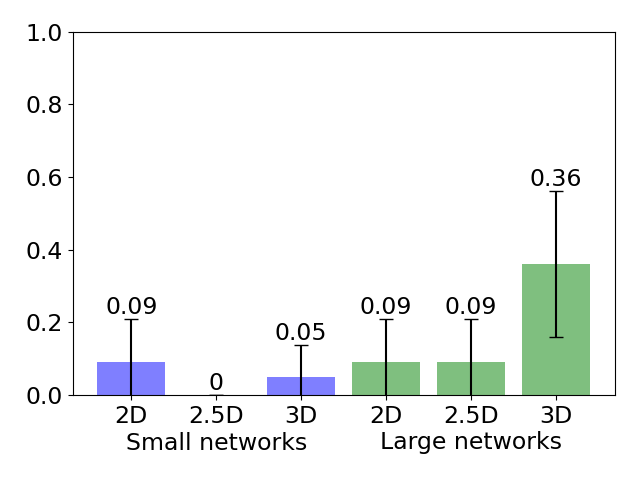}&
\includegraphics[width=0.25\textwidth,trim={20 20 10 10},clip]{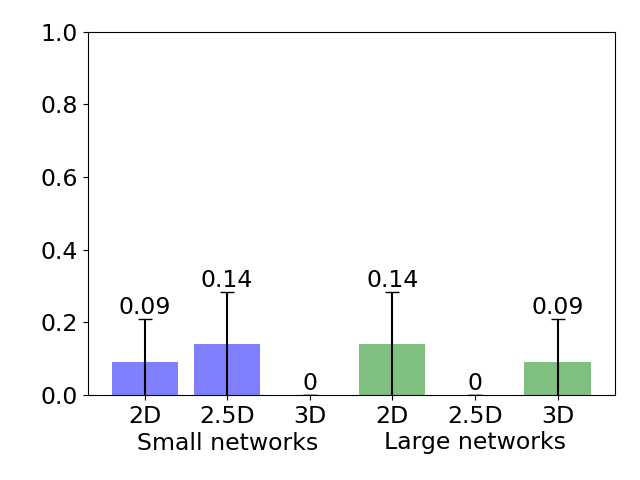}\\
\hline
\multicolumn{3}{|c|}{We found no significant relationship. No test can be computed with T5 due to floor effects.}\\
\hline
\multicolumn{3}{|c|}{Completion Time}\\
\hline
\includegraphics[width=0.25\textwidth,trim={20 20 10 10},clip]{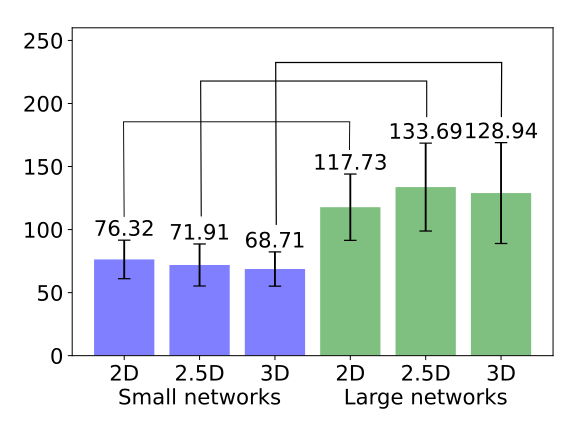}&
\includegraphics[width=0.25\textwidth,trim={20 20 10 10},clip]{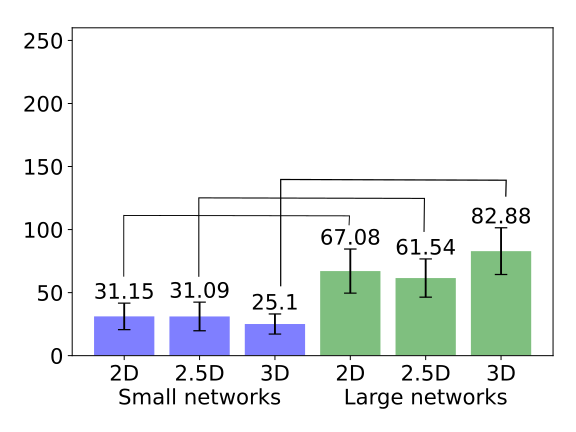}&
\includegraphics[width=0.25\textwidth,trim={20 20 10 10},clip]{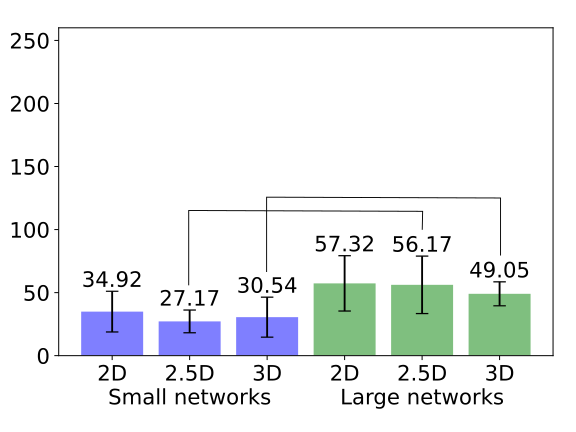}\\
\hline
Significant $p\texttt{-values} \leq 0.002$.&
Significant $p\texttt{-values} < 0.001$.&
Significant $p\texttt{-values} < 0.001$.\\
\hline
\multicolumn{3}{|c|}{Movement performed}\\
\hline
\includegraphics[width=0.25\textwidth,trim={20 20 10 10},clip]{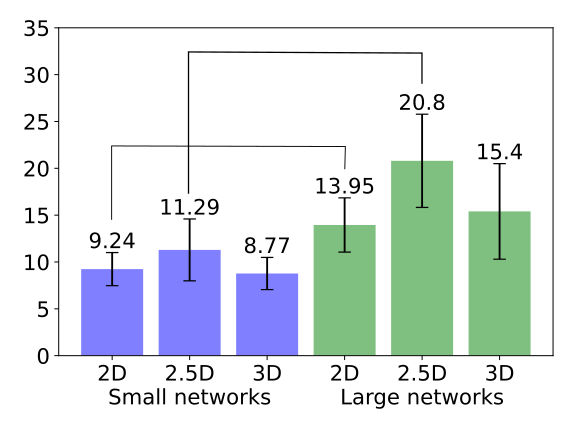}&
\includegraphics[width=0.25\textwidth,trim={20 20 10 10},clip]{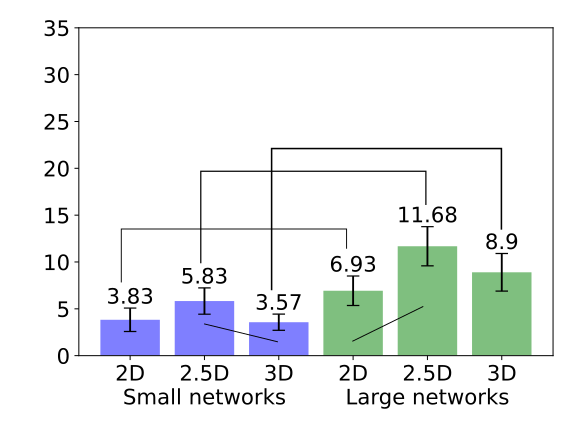}&
\includegraphics[width=0.25\textwidth,trim={20 20 10 10},clip]{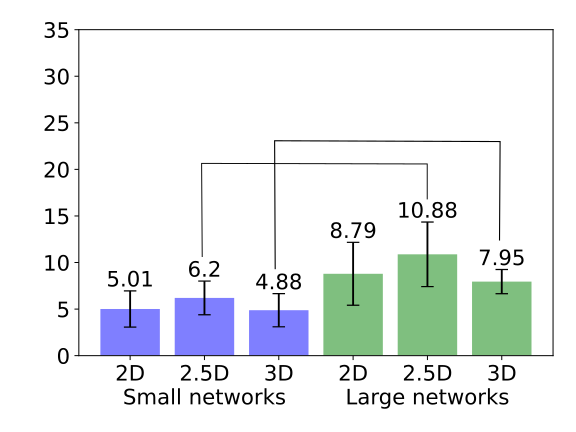}\\
\hline
Significant $p\texttt{-values} \leq 0.001$.&
Significant $p\texttt{-values} \leq 0.004$.&
Significant $p\texttt{-values} < 0.001$.\\
\hline

\end{tabular}

\captionof{figure}{Mean error rate, completion time and movement performed for T4, T5, and T6.}\label{tab:resultsT4-T6}
\end{table*}
}

\subsection{Entity Comparison Across Layers (T3, T4)}

Because T3 (connectivity pattern) is a yes/no question, we first need to confirm that these are not random answers. To this end, we compute another Fisher's exact test on the error rate 
against a Bernoulli random variable with $p=0.5$. All $p\text{-values}>\alpha$, so the test confirms that the data are independent and we do not have random answers.
For small networks, the error rate is the lowest in 3D (0.23) and the highest in 2.5D (0.32). The task completion time is the highest in 3D (106.22s) and the lowest in 2D (80.58s). The participants moved the most in 2.5D (9.86m) and the least in 2D (6.8m). 
For large networks, the error rate is the lowest in 2D (0.18) and the highest in 3D (0.45). The movement performed as well as the task completion time are significantly highest in 2.5D (186.23s, 17.46m) and the lowest in 2D (120.45s, 9.67m).

Task T4 (node degree) has the lowest error rate for small networks in 2.5D (0.09) and the highest in 2D (0.36). Task completion time is homogeneous for small and large networks. Participants moved the most in 2.5D (11.29m) and the least in 3D (8.77m).
For small networks, the error rate is the lowest in 2.5D as well (0.23) and 2D and 3D have the same error rate (0.27).
Task completion time and the movement performed are the highest in 2.5D (133.69s, 20.8m) and the lowest in 2D (117.73s, 15.4m).

\subsection{Layer Comparisons (T5, T6)}
For small networks with T5 (number of edges), there are no errors in 2.5D and the highest error rate is in 2D (0.09). The task completion time is homogeneous between small and large networks. 
For completion time and movement performed, we have significant differences between small and large networks.
Participants performed the most movement in 2.5D (5.83m) and a similar amount in 2D and 3D (3.38m and 3.57m). For large networks, 2D and 2.5D have the lowest error rate (0.09), while 3D has the highest error rate (0.36). The task completion time is the lowest in 2.5D (61.54s) and the highest in 3D (82.88s). Most movement is performed in 2.5D (11.68m) and, the least in 2D (6.93m). 

For small networks with T6 (number of connected components), there are no errors in 3D and 2.5D has the highest error rate (0.14).
For the completion time and movement performed, we have a significant difference between small and large networks when adding a third dimension.
The task completion time is the lowest in 2.5D (27.17s) and the highest in 2D (34.92s). The movement performed is the highest in 2.5D (6.2m) and the lowest in 3D (4.88m). For large networks, there are no errors in 2.5D and the highest is in 2D (0.14). Task completion time is homogeneous. The most movement is performed in 2.5D (10.88m) and, the least in 3D (7.95m).

\subsection{Qualitative Data}\label{sub:qualitative}
After the experiment, participants were asked to assess {their performance and give their opinion} about each task, network complexity, and layer arrangement; see  supplemental materials for more data.

First, the participants were asked to order their favourite layer arrangement for each network complexity.
For small networks, 41\% of the participants preferred 2.5D visualisation followed by 2D (32\%) and 3D (27\%). 
For large networks, 45\% of the participants favoured 3D, followed by 2D (32\%) and 2.5D (23\%).

We also asked participants about the amount of movement they felt they had to perform. Adding a third dimension is physically engaging: 64\% of the participants rated their movement as \cit{high} in 2.5D and 32\% in 3D. Not surprisingly, 73\% of the participants rated their movement as \cit{high} on large networks and 50\% rated their movement as \cit{low} on small networks.

Finally, for each task, layer arrangement and, network complexity, participants were asked to rate the subjective level of difficulty on a 5-point Likert scale.
We computed the ratio of the two top items of the scale 
over the 22 answers.
On small networks, 
the participants responded most positively for 2D over 2.5D/3D except for T2 and T5, where adding the third dimension was appreciated. 
On large networks, 
the participants responded most  positively for 2D for T4 and T5.

\section{Discussion}\label{sec:discussion}
Our results indicate that different layer arrangements allow for certain tasks to be performed more effectively. While we found no clear overall winner, we think that the discussion of our findings contributes to the development of the field, e.g., by leading to the design of more effective MLN visualization techniques.

\subsection{Impact of the Task Taxonomy Category}

\paragraph*{\textbf{Entity Connectivity Across Layers.}}
For T1, the high error rate for 2D is consistent with the presence of edge crossings impacting the readability of the networks~\cite{Purchase1997}. 
Overall, adding a third dimension seems to improve the accuracy without changing the completion time dramatically. The movement performed is significantly higher for 2.5D because the network is about 4 metres long and T1 implies looking in detail at each layer.

The floor effect of T2 for small networks implies that the task is trivial. This is confirmed by the few movements performed and a quick completion time.
For large networks the task is difficult. We have the same order in all charts: 2D is better than 3D which is better than 2.5D.
T2 implies in this case, a high cognitive load because participants have to compare and remember the number of connections for each of the 7 layers (3 participants even ask for something to write down notes). Adding a third dimension, which also increases the cognitive load, does not help for large networks.

\paragraph*{\textbf{Entity Comparison Across Layers.}}
T3 is the longest and the most difficult task of the experiment, despite being a yes/no question. Even if the pattern is small and clear to the participants, we do not notice a learning effect. The complexity of the network has a significant impact on the performance of T3 in 2.5D.

For T4, adding a third dimension helps reduce the error rate, especially in 2.5D. Also, note that the small increase in movement performed for 2.5D has no impact on completion time.
The participants seem to be able to find the right answer on the first return walk of the network.

\paragraph*{\textbf{Layer Comparisons.}}
For both T5 and T6, the task completion time is close between all layer arrangements. 
However, due to the absence of errors in three cases over both tasks, we may also face a floor effect with two too-easy tasks.
The connected components in T6 are most of the time well separated in the layouts, thus making the task an easy one. 
However, we feel that layers can be regarded as visual filters to reduce the impact of visual clutter. The use of the third dimension allows to visually separate layers. 

\subsection{Impact of Network Complexities}
We expected that an increase in the complexity of the networks (from 3 to 7 layers) would result in increased error rates, completion times and movement.
The results, however, are not perfectly clear. 
Of the 18 options, (6 tasks $\times$ 3 layer arrangements), with respect to error rates, the tasks were harder on large networks (12 times) than on small networks (4 times).
Regarding the movement performed, the large networks always needed more movement and the difference was statistically significant in 8 cases.
Completion time was longer for large networks than for small networks.
However, in T1, the highlighted nodes to determine the shortest path are located on different layers and the maximum path length is between 3 and 5 despite the number of layers. Thus, the participants may only consider a subset of the layers to complete the task. Therefore, more layers do not necessarily increase the complexity of this task.

\subsection{Impact of Layer Arrangement}
Our initial hypothesis was that certain MLN tasks can be performed faster and more accurately with 2.5D and 3D representations. In particular, the 2.5D concept corresponds to the frequently used mental model of layers in an MLN \cite{BrandesDwyerSchreiber+2004+11+26,McGee19}. Consequently, we expected that the best results will be associated with  2.5D and 3D representations and that these would also scale better for large networks.

One pattern, that is common across most tasks and complexities, is that the participants tend to move more in 2.5D. This can be explained by the observation that in 2D and 3D the network can be seen in its entirety from one static position, as all the layers are in one plane (2D) or on the surface of a half-sphere (3D). In contrast, in 2.5D the participants need to walk to observe all layers. 
However, more movement does not necessarily imply a higher error rate, although more movement is correlated with longer task completion times.

The qualitative data shows that the 2.5D layout never scores highest for both complexity levels.
One participant stated that \cit{[\ldots 2.5D] is harder to compare because you have to walk so much}. In particular, this is interesting for small networks, because small 2.5D networks are subjectively favoured by the participants. One reason might be that the visualisation itself is visually appealing. Another participant stated that \cit{[\ldots 2.5D] is fun to really walk through the layers}. Another reason might be, that the favourites depend more on the task, than on the layer arrangement, e.g., \cit{The best layout is hard to say, it depends on the task}. Comparing the favourite layer arrangements shows an inverted image between small and large networks. For small networks, the participants' most favoured is 2.5D and the least favoured is 3D. For large networks, the participants' most favoured is 3D and the least favoured is 2.5D. In contrast to the qualitative data about favourite layer arrangements, the qualitative data about the movement the participants performed are congruent with the quantitative data. Not surprisingly, the participants perceived more movement in large networks than in small networks. More movement was also perceived in 2.5D than in the other layer arrangements.

\subsection{Which layer arrangement to use?}
We can now identify groups of tasks that benefit from higher dimensional layer arrangements and tasks for which adding more dimensions is perceptually/cognitively ineffective or harmful. 

\paragraph*{\textbf{T1, T4, T5: Higher-dimensional layer arrangements are better than 2D for small networks.}} While the movement performed during these tasks is higher, it does not result in a higher task completion time. Adding more dimensions seems to help the participants to solve the tasks. In the qualitative data, this is congruent with T5, but T1 and T4 are rated as more feasible in 2D.

\paragraph*{\textbf{T3, T6: 3D representation is better for small networks.}} In T6 the participants searched for connected components inside layers. By design, the connected components cannot overlap. However, in 2D and 2.5D they can be close to each other, and thus hard to distinguish, while in 3D they stay well separated.
2.5D is associated with the highest movement in this group, but this does not result in a higher task completion time. 
The qualitative data is mixed here.

\paragraph*{\textbf{T1, T4, T6: Adding more dimensions yields better results for large networks.}} 
The error rate for these tasks indicates that 2D might not be the best layout.
Moreover, there are little differences in the task completion time. More dimensions do not imply a significantly longer completion time.
In addition, the qualitative data shows no clear trend in this category.

\paragraph*{\textbf{T2, T3: 2D representation is better for large networks.}} 
While performing T2, participants struggled to determine on which layer an edge between layers ends. Some participants even placed their head directly at the beginning of an edge and looked along the line, to see where it ends.
This might explain the statistically significant higher movement and completion time. Participants need to check the visualisation many times before answering the task.
T3 was the hardest task of the evaluation and took the most time. 
Familiarity with the standard 2D representation helped with these two challenging tasks.

\paragraph*{\textbf{T5: Avoid 3D for large networks.}} 
Even if we cannot statistically confirm a significant difference, T5 seems to be harder to perform in the 3D setting. 
3D visualisation  seems to not be suitable to detect layer density. This is confirmed by the qualitative data where 2D received the highest preference.

\subsection{\textbf{Take-Away Messages}}

\paragraph*{\textbf{Mental model for MLNs.}} While the 2.5D layer arrangement matches known mental model of MLNs \cite{brandes, McGee19}, it does not consistently outperform the other options.

\paragraph*{\textbf{Higher dimensions for MLNs.}} For each task taxonomy category, at least one task can be performed better in higher dimensional layer arrangements. Specifically, 3 out of 6 tasks can be performed better in 2.5D or 3D for both small and large networks.
    
\paragraph*{\textbf{Scalability of MLN representations.}} The layer arrangement does not seem to impact scalability, i.e., solving tasks with larger networks tends to require more time and more movement.
     
\paragraph*{\textbf{Movement for MLNs.}} While 2.5D tends to require more movement, there is no correlation with higher error rates.

\section{Limitations}\label{sec:limitations}
While the floor effect for T2 with small networks was feared in favour of a consistent visualisation across all arrangements and tasks, no consistent insights can be acquired in this setting. This might influence the qualitative data for small networks. To overcome any learning effect, we applied a Graeco-Latin square for counterbalancing tasks and some randomisation. But still, more participants should lead to more (statistically) reliable data.

Small and large networks refer to simultaneously increasing the size of the network (number of nodes and edges) and the number of layers (from 3 to 7 layers).
Considering these two aspects separately might have an impact on the participants' performances. 
When creating the networks, we did not consider the number of edge crossings. This also undoubtedly should have an impact on the experimentation.
Furthermore, we did not use a flat 2D MLN visualisation where layers are not visually well separated. This usually results in hairball visualisations. It would be worthy of interest to compare such visualisations against the layer arrangements discussed in this paper.

In VR, the change of perspective on an object is caused by the participant's movement. During the experiment, some participants even knelt down or sat on the floor to have a better look at layers closer to the ground. In our setup networks are realised with fixed height, and so shorter participants could be disadvantaged when looking at layers placed near the top.

For the selection of tasks, we excluded category C (interaction with MLNs and alternating MLNs; see \cref{sec:design}) as this would require a different experimental setup focusing on VR-centred interaction techniques.

Our experiment uses two complexity levels (small and large networks). Adding more levels to the experiment results in another multiplier for the number of tasks participants have to accomplish. This number links directly to the study duration, possible learning effects, and fatigue. 
We chose a trade-off number to allow for comparisons across layer arrangements as this is the main study subject. Moreover, changing the complexity of our networks might influence the outcome of the experiment and may apply to many, but not all real-world scenarios.  
Another influential factor in our study is the layer arrangement.
In 3D, the layers are located on a sphere, which makes it possible to compare the results with the 2D and 2.5D arrangements. 
Many other arrangements could be imagined (e.g., layers could be placed all around the participant) but we did not attempt to explore the complete design space. Different arrangements might yield different results and this seems like relevant future work.

As we wanted to focus on network representation only, an early design decision was to not use interactions. Appropriate interactions can help improve performance and this is worth exploring.

In the design of T1, we controlled the minimum and maximum of the shortest path length, but not the number of layers that might be visited. In T2 the data about the 2.5D layer arrangement might be influenced by the position of the target layer, in case the target layer is at or near the outer layer of the graph.

\section{Conclusion}\label{sec:conclusion}
We explored the use of three visual layer arrangements (2D, 2.5D, and 3D) for the analysis of MLNs of different complexities. A human subject experiment was conducted in virtual reality to evaluate which arrangements and complexities are most effective for a set of six analysis tasks derived from the MLN task taxonomy of McGee et al.~\cite{MLBBook21}. 
The results showed that different tasks can be performed more effectively using different visual settings, and no setting outperforms the others in general. 
The presented study and the related task-to-arrangement metaphor space and its derived recommendations can help to effectively use the different layer arrangements for MLN analysis tasks.

The experiment was conducted in virtual reality to utilise the benefits of stereoscopic view representations of MLNs. It would be interesting to conduct a similar experiment on a 2D screen to investigate the influence of the type of display. In addition, the influence of alternative representations of the 3D layout could be interesting to investigate. A follow-up experiment could also close the gap for task taxonomy category~C (see \cref{sec:design}).

\section*{Supplemental Materials}
All supplemental materials are released under a CC BY 4.0 or GPL license and available at \url{https://doi.org/10.18419/darus-3387}.
We include  
(1) all raw quantitative results (CSV files) and all raw qualitative data,
(2) all generated MLNs (GraphML file format) along with the code of the generator (C++ files using OGDF),
(3) a PDF document containing detailed results (with $p$-values and more charts),
(4) a video presenting the experimentation from a participant's point of view.

\section*{Figure Credits} 
All images are created by the authors.

\acknowledgments{Work on this topic began at Dagstuhl Seminar 21401, ``Visualization of Biological Data.''
We acknowledge funding by DFG, under Germany’s Excellence Strategy – EXC 2117 – 422037984, DFG project ID 251654672 – TRR 161, the ELLIIT environment for strategic research in Sweden, and the National Science Foundation, NSF-CCF-2212130. We also thank Vincent Couallier (U. Bordeaux) for helping with the statistical analysis.}
\bibliographystyle{abbrv-doi-hyperref}

\bibliography{thePaper-crc}

\end{document}